\documentclass[a4paper,11pt]{article}
\usepackage{jcappub} % JCAP style
\usepackage[T1]{fontenc} % if needed
\usepackage{amsmath,amssymb,graphicx}
\usepackage{physics} % for Dirac notation and derivatives
\usepackage{bm}      % bold math symbols
\usepackage{hyperref} % clickable references

\title{Anisotropic Neutrino Emission from Spinning, Moving, and Charged Primordial Black Holes}

\author{Arnab Chaudhuri}

\affiliation{Division of Science, National Astronomical Observatory of Japan, Mitaka, Tokyo 181-8588, Japan.}

\emailAdd{arnab.chaudhuri@nao.ac.jp}

\abstract{
The angular and spectral features of neutrinos emitted from primordial black holes (PBHs) carry key imprints of the black hole’s fundamental properties. This work investigates the directional emission of neutrinos from Kerr-Newman PBHs undergoing Hawking evaporation, accounting for the combined effects of spin, motion, and electric charge. Rotation induces anisotropic fluxes through axisymmetric geometry and spin-dependent greybody factors, while relativistic motion leads to pronounced Doppler beaming along the direction of travel. Electric charge modifies the thermodynamic evolution and suppresses the emission of like-charged particles, altering the overall spectrum and burst duration. The resulting neutrino flux exhibits rich angular structure, energy dependence, and time profiles that vary with PBH parameters. These directional signatures enhance the prospects for detection at current and future neutrino observatories, and offer new multi-messenger probes of PBH populations in the early universe.
}


\begin{document}
\maketitle
\flushbottom

\section{Introduction}

Primordial black holes (PBHs) have long been regarded as unique cosmic laboratories, capable of probing physics beyond the Standard Model, quantum gravity, early-Universe cosmology, and the dark sector~\cite{Zeldovich:1967lct,Hawking:1971ei,Carr:1974nx,Carr:2009jm,Carr:2020xqk,Green:2020jor}. Unlike astrophysical black holes, PBHs can form in the early Universe from large amplitude fluctuations, bubble collisions, cosmic string collapse, or other nonstandard mechanisms~\cite{Garcia-Bellido:1996mdl,Kawasaki:2016pql,Sasaki:2018dmp,Clesse:2015wea}. The mass spectrum of PBHs is model-dependent, spanning from sub-gram to solar-mass and beyond~\cite{Carr:1975qj,Byrnes:2012yx,Green:2016xgy}, and their observable imprints are correspondingly diverse.

Among the most striking signatures of PBHs is their Hawking evaporation~\cite{Hawking:1974rv,Hawking:1975vcx}, a process that leads to the emission of all particle species lighter than the black hole temperature. This emission becomes especially relevant for low-mass PBHs ($M \lesssim 10^{15}~\mathrm{g}$), which can fully evaporate within the age of the Universe, producing photons, neutrinos, gravitons, and possibly new particles such as axions or sterile neutrinos~\cite{MacGibbon:1991vc,Barrau:2003xp,Carr:2005zd,Arbey:2019mbc,Dong:2015yjs,Laha:2019ssq}. In this context, PBHs serve as time-integrated particle accelerators and cosmological beam dumps, offering a unique window into both high-energy and weakly coupled sectors.

Neutrinos are especially powerful messengers in this regard~\cite{Bugaev:2008gw,Lunardini:2019zob,Dasgupta:2020mqg}. Unlike photons or charged particles, they are not deflected or absorbed by cosmic magnetic fields or background plasma, allowing them to travel cosmological distances unimpeded. Evaporating PBHs are expected to emit neutrinos thermally, with a spectrum peaking near the Hawking temperature, which can reach hundreds of MeV or higher depending on the black hole mass. For PBHs with masses near the evaporation threshold ($M \sim 10^9$–$10^{14}$ g), the resulting neutrino signal lies within the sensitivity range of detectors such as Super-Kamiokande, Hyper-Kamiokande, and DUNE~\cite{Super-Kamiokande:2020sgt,Hyper-Kamiokande:2018ofw,DUNE:2020ypp}.

Recent studies have shown that when PBHs are in motion relative to the cosmic frame, the Hawking-emitted neutrinos are not distributed isotropically in the observer frame~\cite{Guo:2023hyp,Coogan:2020tuf,Chaudhuri:2025hen}. Instead, relativistic motion induces Doppler boosting and angular compression, resulting in a collimated neutrino burst aligned with the PBH velocity. Such bursts are not only brighter in a given direction but can also carry information about the spatial distribution and kinematics of the PBH population. These effects open a new observational channel: the directional detection of PBH evaporation via neutrino observatories.

While motion-induced anisotropy has received attention recently, other key features of realistic PBHs remain underexplored. In particular, PBHs may possess non-negligible spin and small electric charge. PBHs can acquire spin through multiple mechanisms: during formation from non-spherical collapse~\cite{Chiba:2017rvs,DeLuca:2019buf,Harada:2017fjm}, from angular momentum in the primordial plasma, or from post-formation accretion. Charged PBHs can arise from local charge asymmetries during formation, and although large charges are disfavored due to discharge processes~\cite{Gibbons:1975kk}, small net charges are not ruled out and can influence the emission rates of charged particles~\cite{Lehmann:2018ejc}.

Theoretical studies of Hawking radiation from rotating (Kerr) and charged (Reissner–Nordström, or more generally Kerr–Newman) black holes show that both spin and charge substantially alter the emission spectrum~\cite{Page:1976df,Page:1976ki,Page:1977um,Duffy:2005ns,Grain:2005my,Konoplya:2019hml}. Rotation introduces angular momentum-dependent greybody factors and modifies the spectral shape via superradiance, especially for bosonic species~\cite{Brito:2015oca}. The angular distribution of emitted particles becomes anisotropic, peaking along the black hole spin axis. When combined with relativistic motion, this yields a doubly distorted angular profile, potentially resulting in highly directional neutrino bursts.

In this work, we develop a unified framework to compute the neutrino emission from PBHs characterized by arbitrary spin, electric charge, and motion. We model the black holes using the Kerr–Newman solution~\cite{Newman:1965my}, with mass $M$, spin parameter $a = J/M$, and charge $Q$, and apply relativistic transformations to analyze the emission in the lab frame. Neutrino emission is computed using spin-dependent greybody factors and spheroidal harmonics~\cite{Teukolsky:1973ha}, with full angular resolution.

Our analysis reveals that the combination of rotation and motion leads to highly directional emission even for moderate values of spin and velocity. The energy spectrum is also modified: relativistic beaming blueshifts the neutrinos in the forward direction, while charge reduces the Hawking temperature and suppresses like-charged particle production. We characterize the resulting angular flux profiles and assess their detectability in existing and upcoming neutrino detectors.

The directional nature of the neutrino flux enhances the detectability of PBHs by increasing the instantaneous flux in a given direction, enabling burst searches in coincidence with gamma-ray or gravitational wave observations. Moreover, directional detection offers a handle on distinguishing PBH evaporation events from the isotropic background of supernova or atmospheric neutrinos~\cite{Lunardini:2006sn}. The morphology of the burst can provide constraints on the PBH velocity distribution and spin orientation, while the total flux encodes information about the PBH mass function and abundance. At cosmological scales, such anisotropic neutrino emission could in principle induce directional fluctuations in the cosmic neutrino background (C$\nu$B), with possible relevance to early-Universe probes such as BBN or the 21cm signal~\cite{Acharya:2020jbv,Mittal:2021egv}. Though speculative, this opens new lines of inquiry into the imprint of PBHs on cosmological observables.

The structure of this paper is as follows. In Sec.~\ref{sec:KNgeometry}, we review the Kerr–Newman geometry and derive the relevant thermodynamic quantities. In Sec.~\ref{sec:spin_anisotropies}, we study the angular anisotropy induced by black hole spin, and in Sec.~\ref{sec:motion}, we analyze the effects of relativistic motion. Sec.~\ref{sec:charge} examines the role of electric charge in modulating emission and evaporation timescales. In Sec.~\ref{sec:Results}, we present the combined angular and spectral results, including lab-frame flux maps for various PBH parameters. Sec.~\ref{sec:detection} discusses detection prospects and signal characteristics at neutrino detectors and in Sec.~\ref{cosmo} we discuss possible cosmological outlook. We conclude in Sec.~\ref{sec:conclusion} with a summary and outlook for future work.

\section{Primordial Black Holes with Spin, Charge, and Motion}
\label{sec:KNgeometry}

We begin by developing the theoretical framework necessary to describe PBHs endowed with spin, electric charge, and relativistic motion. These features, while often neglected in cosmological analyses, can play a crucial role in determining the angular and spectral profiles of Hawking radiation—especially for particles such as neutrinos. To this end, we first review the Kerr–Newman (KN) solution of the Einstein–Maxwell equations and the associated black hole thermodynamics. We then present the formalism for Hawking radiation in this background, including the modifications induced by greybody factors and the effects of angular momentum and electric charge.

\subsection{Kerr–Newman Geometry and Thermodynamics}
\label{subsec:KN_thermo}

The Kerr–Newman metric represents the most general asymptotically flat, stationary solution to the Einstein–Maxwell equations in four spacetime dimensions, describing a rotating, charged black hole characterized by three parameters: the mass $M$, the specific angular momentum $a \equiv J/M$, and the electric charge $Q$~\cite{Newman:1965my, Misner:1973zz}. This solution plays a central role in our analysis as it captures the relevant spacetime geometry around spinning and charged PBHs.

In Boyer–Lindquist coordinates $(t, r, \theta, \phi)$, the Kerr–Newman metric takes the form
\begin{equation}
\begin{aligned}
    ds^2 &= -\left(1 - \frac{2Mr - Q^2}{\Sigma} \right) dt^2 
    - \frac{2a\sin^2\theta (2Mr - Q^2)}{\Sigma} dt\, d\phi 
    + \frac{\Sigma}{\Delta} dr^2 
    + \Sigma\, d\theta^2 \\
    &\quad + \sin^2\theta \left( r^2 + a^2 + \frac{(2Mr - Q^2)a^2\sin^2\theta}{\Sigma} \right) d\phi^2,
\end{aligned}
\end{equation}
with the functions $\Sigma$ and $\Delta$ defined by
\begin{align}
    \Sigma(r, \theta) &= r^2 + a^2 \cos^2\theta, \\
    \Delta(r) &= r^2 - 2 M r + a^2 + Q^2.
\end{align}

The locations of the outer and inner horizons are determined by the roots of \( \Delta = 0 \), yielding:
\begin{equation}
    r_\pm = M \pm \sqrt{M^2 - a^2 - Q^2},
\end{equation}
where \( r_+ \) corresponds to the event horizon and \( r_- \) to the Cauchy (inner) horizon.
The condition for the existence of a regular event horizon—avoiding a naked singularity—is the so-called extremality bound:
\begin{equation}
    a^2 + Q^2 \leq M^2.
\end{equation}
The extremal case, where equality is saturated, corresponds to a black hole with vanishing surface gravity and, hence, zero Hawking temperature.

Thermodynamically, the surface gravity $\kappa$ is related to the Hawking temperature by $T_H = \kappa/(2\pi)$. For the Kerr–Newman geometry, this yields
\begin{equation}
    T_H = \frac{r_+ - r_-}{4\pi (r_+^2 + a^2)} = \frac{\sqrt{M^2 - a^2 - Q^2}}{2\pi \left( r_+^2 + a^2 \right)}.
\end{equation}
This expression highlights how both spin and charge reduce the temperature, prolonging the black hole’s lifetime. For primordial black holes in the early Universe, this suppression can significantly affect the timing and nature of their evaporation.

The area of the event horizon is given by
\begin{equation}
    A_H = 4\pi (r_+^2 + a^2),
\end{equation}
from which the black hole entropy follows via the Bekenstein–Hawking formula,
\begin{equation}
    S_{BH} = \frac{A_H}{4} = \pi (r_+^2 + a^2),
\end{equation}
in natural units ($G = \hbar = c = k_B = 1$).

The first law of black hole thermodynamics for Kerr–Newman black holes generalizes the energy conservation relation:
\begin{equation}
    dM = T_H\, dS + \Omega_H\, dJ + \Phi_H\, dQ,
\end{equation}
where $\Omega_H$ and $\Phi_H$ denote the angular velocity and electric potential at the horizon, respectively:
\begin{align}
    \Omega_H &= \frac{a}{r_+^2 + a^2}, \\
    \Phi_H &= \frac{Q r_+}{r_+^2 + a^2}.
\end{align}
These quantities influence the energy spectrum and angular distribution of particles emitted during Hawking evaporation.

\subsection{Hawking Radiation from Kerr–Newman PBHs}
\label{subsec:Hawking_KN}

Hawking radiation is a quantum process through which black holes emit particles thermally, with a characteristic spectrum determined by the black hole’s temperature and the properties of the emitted particles. For Kerr–Newman black holes—characterized by mass, spin, and electric charge—the emission spectrum is considerably more complex due to the interplay of rotation, charge, and the horizon’s geometry.

The differential emission rate per unit time and energy for particles of spin $s$ can be expressed as
\begin{equation}
\label{eq:hawking_spectrum}
    \frac{d^2N_s}{dt\, d\omega} = \sum_{\ell = s}^\infty \sum_{m = -\ell}^\ell \frac{1}{2\pi} \frac{\Gamma_{s\ell m}(\omega)}{\exp\left[\frac{\omega - m \Omega_H - q \Phi_H}{T_H}\right] \pm 1},
\end{equation}
where:
\begin{itemize}
    \item $\omega$ is the energy of the emitted particle,
    \item $q$ denotes its electric charge,
    \item $\Gamma_{s\ell m}(\omega)$ represents the greybody factor corresponding to the mode $(\ell,m)$,
    \item and the $+$ ($-$) sign corresponds to fermions (bosons).
\end{itemize}

The greybody factors arise from the probability that emitted particles overcome the curved spacetime potential barrier surrounding the black hole. These factors depend sensitively on the black hole parameters and must be computed by numerically solving the Teukolsky equation with appropriate boundary conditions for Kerr–Newman spacetimes~\cite{Teukolsky:1973ha, Duffy:2005ns}. A comprehensive discussion of the greybody factor calculation is provided in Appendix~\ref{app:greybody}.

The denominator in Eq.~\eqref{eq:hawking_spectrum} contains the generalized Boltzmann factor modified by chemical potentials associated with the black hole’s rotation and charge. The term $(\omega - m \Omega_H - q \Phi_H)$ effectively shifts the energy of emitted particles due to the horizon’s angular velocity $\Omega_H$ and electric potential $\Phi_H$. For bosonic particles, this shift can render the effective energy negative, leading to \emph{superradiant amplification} of certain modes. In contrast, fermions, such as neutrinos, do not exhibit superradiance because of the Pauli exclusion principle, although their emission spectra are still affected by frame dragging and electromagnetic interactions.

Electric charge influences emission rates by suppressing particles carrying the same sign of charge as the black hole and enhancing oppositely charged species. Since neutrinos are electrically neutral, the charge $Q$ primarily affects their emission through modifications to the black hole temperature and metric background.

Integrating over energy, the total emission rates for particle number and energy are given by
\begin{align}
    \frac{dN_s}{dt} &= \int_0^\infty d\omega \sum_{\ell = s}^\infty \sum_{m = -\ell}^\ell \frac{1}{2\pi} \frac{\Gamma_{s\ell m}(\omega)}{\exp\left[\frac{\omega - m \Omega_H - q \Phi_H}{T_H}\right] \pm 1}, \\
    \frac{dE_s}{dt} &= \int_0^\infty d\omega \, \omega \sum_{\ell = s}^\infty \sum_{m = -\ell}^\ell \frac{1}{2\pi} \frac{\Gamma_{s\ell m}(\omega)}{\exp\left[\frac{\omega - m \Omega_H - q \Phi_H}{T_H}\right] \pm 1}.
\end{align}
These rates govern the evolution of the black hole’s mass, angular momentum, and charge, as well as the spectra of emitted particles.

For primordial black holes formed in the early Universe, even modest spin and charge can substantially alter the evaporation process compared to the Schwarzschild case. These modifications become particularly significant when investigating directional signals, such as anisotropic neutrino bursts, emitted during the late stages of evaporation.

In the following section, we explore how spin, charge, and bulk motion collectively shape the angular and spectral characteristics of neutrino emission, highlighting their potential as unique probes of primordial black hole properties.

\section{Spin-Induced Angular Anisotropies}
\label{sec:spin_anisotropies}

The rotation of Kerr–Newman primordial black holes induces pronounced angular anisotropies in the emitted particle flux, in stark contrast to the isotropic emission expected from non-rotating Schwarzschild black holes. This anisotropy arises primarily from frame-dragging effects and the nontrivial angular structure of wave modes propagating in a rotating spacetime background.

\subsection{Spin-Weighted Spheroidal Harmonics and Angular Decomposition}

The angular distribution of emitted particles is governed by the spin-weighted spheroidal harmonics \( S_{\ell m}^{(a\omega)}(\theta, \phi) \), which are eigenfunctions of the angular Teukolsky equation~\cite{Teukolsky:1973ha,Seidel:1988ue}:

\begin{equation}
    \left[
    \frac{1}{\sin\theta} \frac{d}{d\theta} \left(\sin\theta \frac{d}{d\theta} \right) + (a \omega)^2 \cos^2\theta - \frac{m^2}{\sin^2\theta} - 2 a \omega s \cos\theta - s^2 \cot^2\theta + E_{\ell m s}
    \right] S_{\ell m}^{(a\omega)}(\theta) = 0,
\end{equation}
where \(s\) is the spin-weight of the particle field (e.g., \(s = \pm \frac{1}{2}\) for neutrinos), \(a\) is the black hole spin parameter, \(\omega\) the mode frequency, and \(E_{\ell m s}\) the angular eigenvalue.

These harmonics form an orthonormal basis on the sphere:
\begin{equation}
    \int_0^\pi \left| S_{\ell m}^{(a\omega)}(\theta) \right|^2 \sin\theta\, d\theta = 1,
\end{equation}
and reduce to the standard spin-weighted spherical harmonics \( {}_sY_{\ell m}(\theta, \phi) \) in the non-rotating limit \(a\omega \to 0\). In this limit, the angular dependence of the emitted radiation is effectively captured by \(|{}_sY_{\ell m}(\theta, \phi)|^2\), which describes the distribution of particle flux as a function of polar angle for given spin weight \(s\), total angular momentum \(\ell\), and azimuthal number \(m\). To illustrate this structure and its relevance for anisotropic Hawking radiation, we consider the special case \(\phi = 0\) and examine \(|{}_sY_{\ell m}(\theta, 0)|^2\). This serves as a useful reference profile for understanding the deformation induced by black hole spin, and highlights the angular anisotropies present even in the slowly rotating regime.

\begin{figure}[t]
    \centering
    \includegraphics[width=0.6\textwidth]{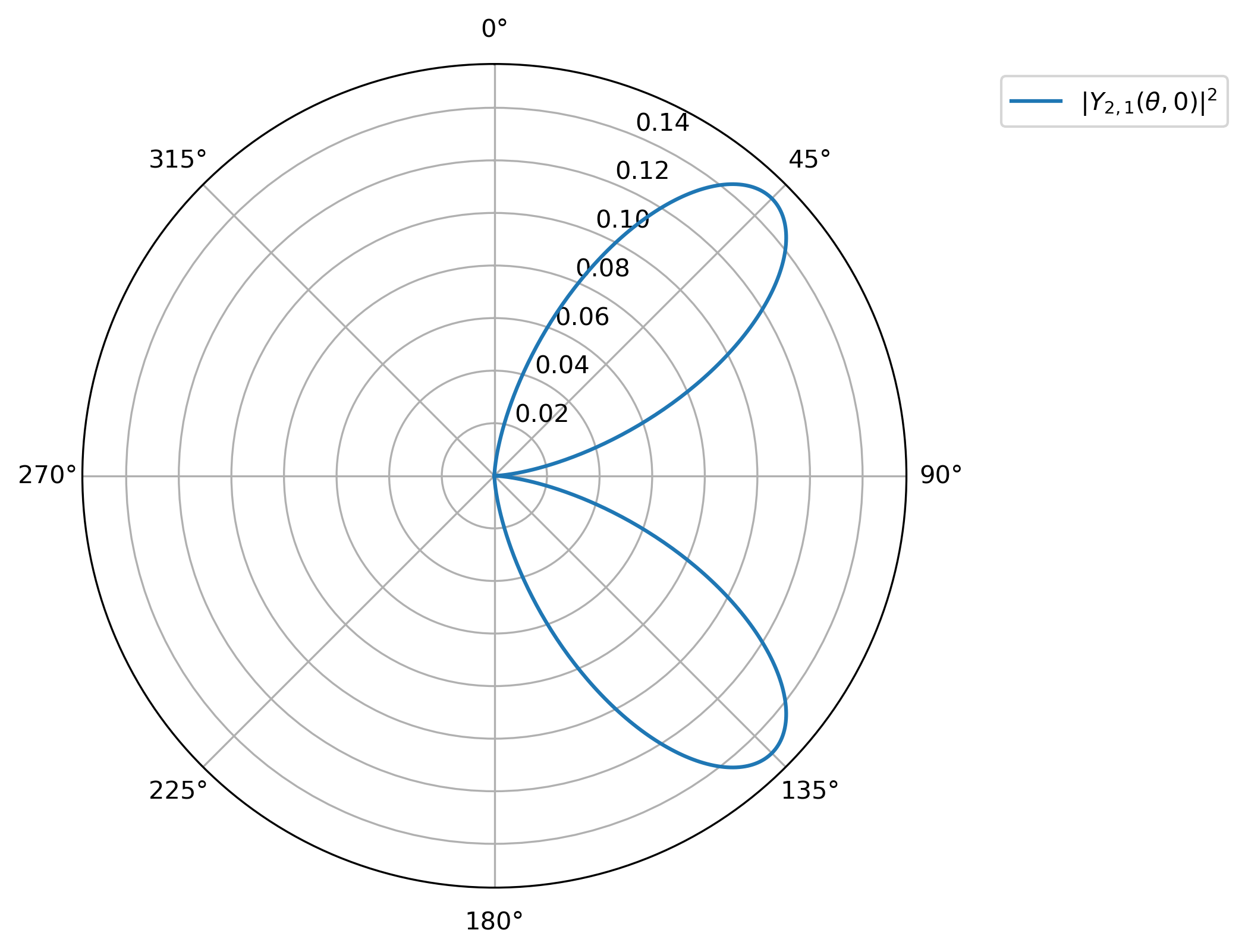}
    \caption{Angular intensity pattern \(|Y_{\ell m}(\theta, 0)|^2\) for the \(\ell=2, m=1\) spherical harmonic mode, illustrating the typical angular dependence of neutrino emission modes in the spin-0 limit. This provides a baseline for understanding the more complex angular profiles induced by rotation and spin-weight effects.}
    \label{fig:sph_harm_theta}
\end{figure}

Figure~\ref{fig:sph_harm_theta} displays the angular intensity \(|Y_{\ell m}(\theta, 0)|^2\) for the \(\ell=2, m=1\) mode. While this corresponds to the standard scalar spherical harmonic (\(s=0\)) and neglects the spin-weight of emitted particles, it nonetheless captures key qualitative features of angular anisotropy. In realistic Hawking emission of fermions or vector bosons, spin-weighted harmonics \({}_sY_{\ell m}(\theta, \phi)\) — and their spheroidal generalizations in the Kerr geometry — become relevant, leading to further deformation and spin-axis alignment of the radiation pattern.

\subsection{Angular Distribution of the Neutrino Flux}

The differential neutrino emission rate per unit time, energy, and solid angle \(d\Omega = \sin\theta d\theta d\phi\) can be expressed as
\begin{equation}
    \frac{d^3 N_{\nu}}{dt\, d\omega\, d\Omega} = \sum_{\ell = s}^\infty \sum_{m = -\ell}^\ell \frac{1}{2\pi} \frac{\Gamma_{s\ell m}(\omega)}{\exp\left[\frac{\omega - m \Omega_H}{T_H}\right] + 1} \left| S_{\ell m}^{(a\omega)}(\theta) \right|^2,
\end{equation}
where the charge term is omitted because neutrinos are electrically neutral.

The emission is azimuthally symmetric about the black hole’s spin axis, rendering the \(\phi\)-dependence trivial for the emission intensity.

\subsection{Mode Contributions and Spin-Dependence}

Each mode \((\ell, m)\) contributes a distinct angular pattern modulated by the greybody factor \(\Gamma_{s\ell m}(\omega)\). Modes with positive \(m\) are typically enhanced due to the co-rotation of the particle wavefunction with the black hole horizon, reflecting frame-dragging effects.

Integrating over energy, the total angular flux distribution is
\begin{equation}
    \frac{dN_{\nu}}{dt\, d\Omega} = \int_0^\infty d\omega \sum_{\ell, m} \frac{1}{2\pi} \frac{\Gamma_{s\ell m}(\omega)}{\exp\left[\frac{\omega - m \Omega_H}{T_H}\right] + 1} \left| S_{\ell m}^{(a\omega)}(\theta) \right|^2.
\end{equation}

\subsection{Observer Orientation and Observable Angular Profiles}

An observer situated at an angle \(\theta_{\text{obs}}\) relative to the PBH spin axis measures an intensity proportional to
\begin{equation}
    I(\theta_\text{obs}) = \int d\omega \sum_{\ell, m} \frac{\Gamma_{s\ell m}(\omega)}{2\pi \left[\exp\left(\frac{\omega - m \Omega_H}{T_H}\right) + 1\right]} \left| S_{\ell m}^{(a\omega)}(\theta_\text{obs}) \right|^2,
\end{equation}
providing a direct probe of both the spin magnitude and orientation.

\subsection{Superradiance and Its Role}

For bosonic fields, modes satisfying the superradiance condition \(\omega < m \Omega_H\) undergo amplification~\cite{Teukolsky:1974yv,Cardoso:2013krh}. Neutrinos, as fermions, do not experience superradiance due to the Pauli exclusion principle~\cite{Unruh:1974bw}, but the emission remains anisotropically enhanced along co-rotating modes.

Superradiant amplification significantly affects bosonic emission angular patterns, often increasing equatorial emission, which is an important consideration in multi-messenger interpretations of PBH evaporation signals.

\section{Relativistic Motion and Doppler Beaming}
\label{sec:motion}

The motion of a PBH with velocity $\mathbf{v}$ relative to the observer further modifies the Hawking emission spectrum via relativistic Doppler effects. 
Lorentz boosting transforms both the energy and the emission direction of the particles, leading to a forward collimation of the flux---commonly referred to as \emph{Doppler beaming}. 
When combined with spin-induced anisotropies, this can produce sharply directional bursts of neutrinos.

\subsection{Boosted Hawking Spectrum}
\label{subsec:boosted-spectrum}
In the rest frame of a rotating black hole, the angular distribution of Hawking radiation is not described by the usual spherical harmonics \( Y_{\ell m}(\theta,\phi) \), but rather by the spin-weighted spheroidal harmonics \( {}_sS_{\ell m}(\theta; a\omega) \). These arise from separating variables in the Teukolsky equation for fields of spin \(s\) propagating in the Kerr background \cite{Teukolsky:1974yv}. For massless neutrinos, the relevant spin weight is \( s = \pm \tfrac{1}{2} \), and the angular profile of the emission spectrum in the rest frame is governed by \( |{}_sS_{\ell m}(\theta; a\omega)|^2 \).

These functions encode the coupling between the particle's spin and the spacetime's rotation. In the limit \( a\omega \to 0 \), they reduce to the familiar spin-weighted spherical harmonics \( {}_sY_{\ell m} \). For scalar emission (\(s=0\)), this further reduces to \( Y_{\ell m} \), but for spinor and vector particles, the angular distributions differ significantly, especially at high frequencies and near-extremal spins.

In this section, we investigate how this rest-frame spectrum transforms under a relativistic boost. The motion of the black hole introduces anisotropic Doppler shifts, compressing the emission into a forward cone along the direction of motion. The full boosted spectrum thus depends on both the intrinsic angular structure encoded in \( {}_sS_{\ell m} \) and the kinematics of the Lorentz transformation.

Let $(E',\theta',\phi')$ denote the neutrino energy and emission angles in the PBH rest frame. 
If the PBH moves with velocity $\beta \equiv v/c$ along the $+z$ axis in the observer frame, the Lorentz transformation gives
\begin{align}
    E &= \gamma E'(1 + \beta \cos\theta') \,,
    \label{eq:energy-boost}
    \\
    \cos\theta &= \frac{\cos\theta' + \beta}{1 + \beta \cos\theta'} \,,
    \label{eq:angle-boost}
\end{align}
where $\gamma = (1 - \beta^2)^{-1/2}$.

The differential emission rate in the rest frame is
\begin{equation}
    \frac{d^3N'}{dt'\,dE'\,d\Omega'} 
    = \frac{\Gamma_{\ell m}(E',a_*)}{2\pi}
    \frac{1}{\exp\!\left[E'/T_{\rm H}\right] + 1}
    \, |Y_{\ell m}(\theta',0)|^2 ,
    \label{eq:rest-frame-spectrum}
\end{equation}
where $T_{\rm H}$ is the Hawking temperature and $\Gamma_{\ell m}$ is the greybody factor.

Under a Lorentz boost, phase space transforms as
\begin{equation}
    \frac{d^3N}{dt\,dE\,d\Omega} 
    = \frac{d^3N'}{dt'\,dE'\,d\Omega'}
      \frac{dE'}{dE} \frac{d\Omega'}{d\Omega} \frac{dt'}{dt} .
\end{equation}
Using Eqs.~\eqref{eq:energy-boost}--\eqref{eq:angle-boost}, and noting that $dt = \gamma\,dt'$ and $dE'/dE = [\gamma(1+\beta\cos\theta')]^{-1}$, we find
\begin{equation}
    \frac{d^3N}{dt\,dE\,d\Omega} 
    = \frac{1}{\gamma^2(1+\beta\cos\theta')^2}
      \frac{d^3N'}{dt'\,dE'\,d\Omega'} .
    \label{eq:boosted-spectrum}
\end{equation}
Substituting Eq.~\eqref{eq:rest-frame-spectrum} gives the full boosted distribution.

The factor $(1+\beta\cos\theta')^{-2}$ represents the \emph{Doppler beaming}, compressing the emission into a cone of angular width $\sim \gamma^{-1}$ around the direction of motion.

\subsection{Combining Spin and Motion}
\label{subsec:spin-motion}

For a spinning and moving PBH, the observed angular distribution is shaped by both the spin-induced anisotropy $|Y_{\ell m}(\theta',0)|^2$ in the rest frame and the Doppler beaming due to motion.

Let $\hat{\mathbf{s}}$ and $\hat{\mathbf{v}}$ denote the unit vectors along the spin axis and velocity vector, respectively, with $\Theta_{\rm sv}$ the angle between them. 
The rest-frame emission pattern is aligned with $\hat{\mathbf{s}}$, while the boost-induced compression is along $\hat{\mathbf{v}}$. 
The combined transformation requires a rotation of the spherical harmonic pattern before applying Eqs.~\eqref{eq:energy-boost}--\eqref{eq:angle-boost}.

We distinguish three special cases:
\begin{itemize}
    \item \textbf{Aligned} ($\Theta_{\rm sv}=0$): Spin anisotropy and Doppler beaming reinforce each other, producing the most strongly collimated neutrino burst.
    \item \textbf{Anti-aligned} ($\Theta_{\rm sv}=\pi$): The spin anisotropy enhances emission opposite to the boost direction, partially counteracting the beaming and broadening the observed cone.
    \item \textbf{Perpendicular} ($\Theta_{\rm sv}=\pi/2$): The beaming cone cuts across the spin-enhanced emission lobes, yielding an asymmetric two-lobed pattern in the observer frame.
\end{itemize}

In the general case, the observed differential flux is
\begin{equation}
    \frac{d^3N}{dt\,dE\,d\Omega} 
    = \frac{1}{\gamma^2(1+\beta\cos\theta')^2}
    \frac{\Gamma_{\ell m}(E',a_*)}{2\pi}
    \frac{1}{\exp(E'/T_{\rm H}) + 1}
    \, \left| Y_{\ell m}(\theta'_{\rm rot},\phi'_{\rm rot}) \right|^2 ,
    \label{eq:combined-boost}
\end{equation}
where $(\theta'_{\rm rot},\phi'_{\rm rot})$ are the rest-frame angles after rotating the spin axis by $\Theta_{\rm sv}$ into the velocity direction.

This framework allows us to compute the full anisotropic and energy-dependent neutrino flux for arbitrary spin and motion configurations, which we will apply in Sec.~\ref{sec:Results} to quantify detection prospects.

\section{Charge Effects on Emission and Evolution}
\label{sec:charge}

In realistic scenarios, primordial black holes (PBHs) may carry a small net electric charge, either inherited from early Universe plasma asymmetries or acquired through selective accretion and emission processes. Although astrophysical black holes are expected to be nearly neutral due to efficient charge neutralization by surrounding plasma, even a small amount of charge can significantly influence the thermodynamics and Hawking emission spectra of microscopic PBHs. In this section, we analyze how charge modifies the black hole temperature, affects the emission rates of charged versus neutral particles, and alters the evaporation dynamics, with a focus on its implications for neutrino production.

\subsection{Hawking Temperature and Charge-Induced Suppression of Emission}
\label{sec:charge-temp}

The Hawking temperature of a general Kerr--Newman black hole, which includes both electric charge \( Q \) and spin \( a = J/M \), is given by
\begin{equation}
    T_H = \frac{1}{2\pi} \cdot \frac{\sqrt{M^2 - a^2 - Q^2}}{r_+^2 + a^2}, \qquad
    r_+ = M + \sqrt{M^2 - a^2 - Q^2},
    \label{eq:KN-temperature}
\end{equation}
where all quantities are expressed in natural units. This expression reduces to the Schwarzschild, Reissner--Nordström, or Kerr temperatures in the respective limits \( Q \to 0 \), \( a \to 0 \), or both.

The temperature vanishes in the extremal limit \( M^2 = a^2 + Q^2 \), halting evaporation. Thus, charge and spin both act to suppress Hawking radiation. In particular, increasing \( Q \) or \( a \) lowers \( T_H \), delaying the black hole's mass loss and lengthening its lifetime.

Charge further modifies the greybody factors by altering the effective potential barrier for emitted particles. Due to Coulomb repulsion, emission of like-charged species is suppressed while oppositely charged particles are emitted more efficiently~\cite{Gibbons:1975kk, Page:1977um}. This leads to preferential discharge, dynamically driving the black hole toward a neutral state. 

Although neutrinos are electrically neutral, the overall energy budget and spectrum of all emitted species are modified by the charge and spin. These factors indirectly impact neutrino observables by altering the PBH's temperature evolution and total flux.

In cosmological settings, environmental conditions or limited particle content could inhibit full discharge, allowing residual charge to persist and influence late-time emission.

\subsection{Impact on Neutrino Emission}
\label{sec:charge-neutrinos}

Neutrinos, being electrically neutral, do not couple directly to the black hole’s charge. Nevertheless, the presence of \( Q \) still affects the neutrino spectrum indirectly in two ways.

First, the reduced Hawking temperature lowers the overall energy scale of emitted radiation, including neutrinos. This leads to a softer neutrino spectrum and a lower total flux compared to an uncharged black hole of the same mass.

Second, the greybody factors for spin-1/2 fields are modified by the charged geometry. The effective potential experienced by fermions depends on the spacetime curvature, which in turn depends on \( Q \)~\cite{Page:1976ki, Konoplya:2010kv}. While the Coulomb potential does not directly affect neutrinos, the change in metric functions leads to subtle differences in the transmission probability, especially at low energies. Consequently, the angular and spectral distributions of neutrinos are distorted relative to the Schwarzschild case.

Although the suppression of neutrino flux is not as dramatic as for charged particles, the cumulative effect on the total energy budget and spectral profile is non-negligible when \( Q/M \gtrsim \mathcal{O}(0.1) \). These effects may influence the detectability of neutrino bursts from evaporating PBHs, particularly in models predicting extended lifetimes or delayed evaporation due to charge retention.

\subsection{Evaporation Timescale and Final Stages}
\label{sec:charge-lifetime}

The charge of a PBH also plays a central role in determining its lifetime. Since the Hawking temperature is reduced by \( Q \), the rate of energy loss is correspondingly suppressed. The evaporation timescale for a charged black hole is thus longer than its neutral counterpart:
\begin{equation}
    \tau_{\text{evap}}(Q) > \tau_{\text{evap}}(Q = 0).
\end{equation}
If the PBH can efficiently radiate charged particles, it will discharge early and subsequently follow a neutral evaporation trajectory. However, in scenarios where charged particle emission is suppressed — e.g., due to Pauli blocking, high plasma densities, or asymmetries in available charge carriers — the PBH may retain a nonzero charge for a significant fraction of its evolution. In such cases, the total evaporation process can be prolonged, delaying the final burst and altering the time-dependent neutrino flux. 

\subsection{Charged PBHs in the Early Universe}

While astrophysical black holes are expected to be nearly neutral due to rapid accretion of opposite charges, PBHs formed in the early Universe may acquire net charge via statistical fluctuations, charge asymmetries in the ambient plasma, or coupling to exotic U(1) gauge fields. Models involving charged PBHs as dark matter candidates or as exotic remnants of inflationary dynamics have been proposed~\cite{Carr:2020gox, Aalsma:2021bit}, and the charge may serve as a useful handle for distinguishing PBHs from other compact objects.

In our context, charge introduces an additional source of anisotropy in the energy flux and modifies the spectral shape of emitted neutrinos, especially during the final stages of evaporation. A careful accounting of these effects is necessary when predicting observables such as neutrino fluence, angular distributions, and burst durations.

\section{Neutrino Angular and Energy Spectra}
\label{sec:Results}

In this section, we analyze the final observable neutrino flux emitted by spinning, charged, and moving PBHs. We combine all the effects discussed in the previous sections—spin-induced angular anisotropy, relativistic Doppler beaming, and charge-suppressed Hawking emission—to derive the full lab-frame angular and energy distributions. These spectra are crucial for evaluating detectability in terrestrial or space-based neutrino observatories.

\subsection{Lab-Frame Angular Flux Distribution}
\label{subsec:angular_flux}

The angular distribution of neutrino emission in the PBH rest frame is shaped by spin-induced anisotropy, characterized by the spin-weighted spheroidal harmonics \( {}_sS_{\ell m}(\theta; a\omega) \), and by the greybody transmission probabilities \( \Gamma_s(\omega, a, Q) \). When the PBH is moving with velocity \( \vec{\beta} \), the angular pattern is Lorentz-boosted, leading to compression of the flux into a narrow cone along the direction of motion. 

To compute the lab-frame angular distribution, we Lorentz-transform the differential emission rate:
\begin{align}
\left. \frac{d^3N}{dtdE d\Omega} \right|_{\text{lab}} = \sum_{\ell, m} \frac{\Gamma_{\ell m s}(E', a, Q)}{2\pi} \cdot \left|{}_sS_{\ell m}(\theta', aE')\right|^2 \cdot \left[ \frac{1}{\exp(E'/T_H) \pm 1} \right] \cdot \left( \frac{dE'}{dE} \right) \cdot \left( \frac{d\Omega'}{d\Omega} \right),
\end{align}
where \( E' \) and \( \theta' \) are the energy and angle in the PBH rest frame, related to the lab-frame via:
\begin{align}
E' = \gamma E (1 - \beta \cos\theta), \quad \cos\theta' = \frac{\cos\theta - \beta}{1 - \beta \cos\theta},
\end{align}
with \( \gamma = 1/\sqrt{1 - \beta^2} \).

\vspace{1em}
\begin{figure}[h!]
\centering
\includegraphics[width=0.75\textwidth]{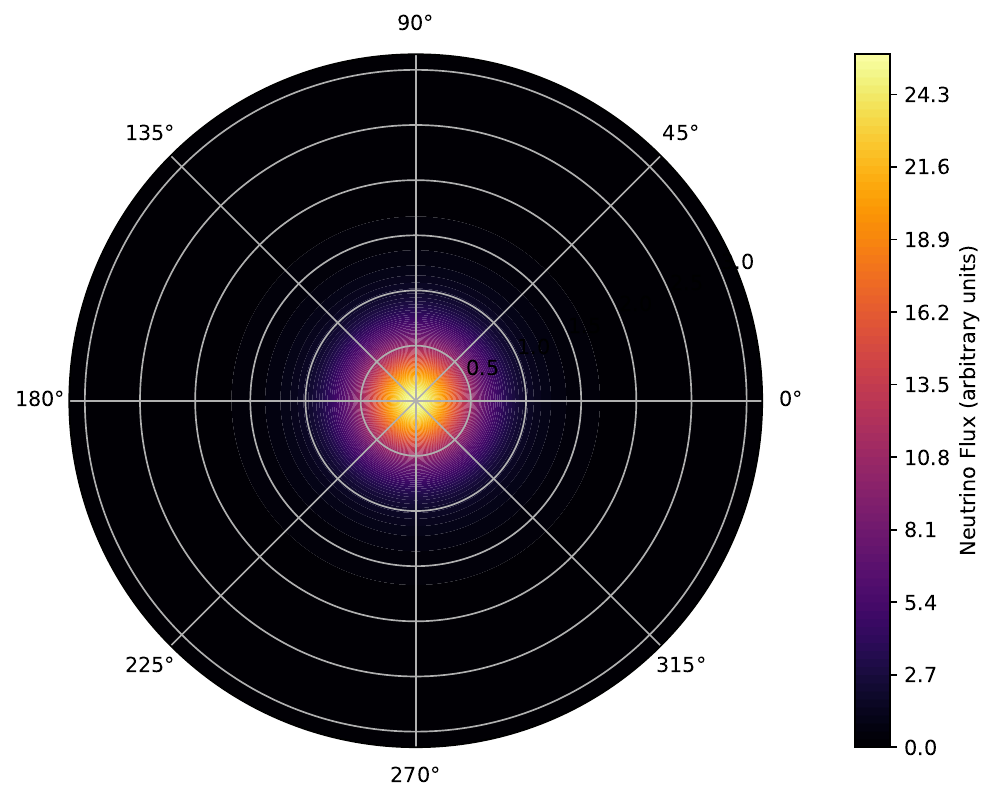}
\caption{Lab-frame angular distribution \( dN/d\Omega \) for neutrinos emitted from a Kerr-Newman PBH with dimensionless spin \( a = 0.8 \), velocity \( \beta = 0.9 \), and charge \( Q = 0.3M \). The beaming effect sharply enhances forward emission along the PBH velocity vector.}
\label{fig:angular_flux}
\end{figure}
\vspace{1em}

Figure~\ref{fig:angular_flux} displays the lab-frame angular distribution of the neutrino flux emitted by a boosted, spinning, and charged PBH. For this plot, we consider a benchmark PBH mass of \( M_{BH} = 10^{12}~\text{g} \), corresponding to a Hawking temperature of \( T_H \approx 1~\text{GeV} \) in the uncharged, non-rotating limit. This choice ensures that neutrinos are among the dominant species emitted during evaporation. The flux is normalized to its maximum value to highlight the anisotropic structure, with the absolute amplitude scaling with \( M^{-2} \) via the temperature dependence of the emission rate. 
The map is plotted in polar coordinates, with the radial coordinate representing the polar angle \( \theta \) (measured from the boost direction) and the azimuthal coordinate \( \phi \). The flux is computed by combining the relativistic Doppler beaming factor with a spin-induced anisotropy model, representative of the angular distribution expected from Kerr-Newman PBHs.

As shown in the figure, the flux exhibits a pronounced forward-peaked structure centered around \( \theta = 0 \), corresponding to the direction of the PBH's motion. This collimation is a direct result of relativistic Doppler boosting, which compresses the emission cone along the velocity axis and enhances the intensity in that direction by a factor approximately proportional to \( (1 - \beta \cos\theta)^{-3} \). For the representative boost value \( \beta = 0.7 \), this enhancement is substantial, causing most of the neutrino flux to be emitted within a narrow cone of half-angle \( \theta_{\text{beam}} \sim \sqrt{1 - \beta}/\gamma \), consistent with special relativistic expectations.

Superimposed on the Doppler enhancement is the angular modulation induced by the black hole's spin. In the rest frame, the emission spectrum of massless particles like neutrinos from a Kerr black hole is not isotropic but follows the profile governed by spin-weighted spheroidal harmonics \( {}_sS_{\ell m}(\theta; a\omega) \). This leads to preferential emission in the equatorial plane for moderate spins and along the poles for extremal configurations. When boosted, these features are distorted, with equatorial lobes being swept forward into off-axis peaks. The resulting lab-frame pattern thus reflects the interplay between angular momentum-induced asymmetry and velocity-induced compression.

Furthermore, while neutrinos are electrically neutral, the presence of charge \( Q \) in the PBH affects the temperature \( T_H \) and consequently the overall emission rate. Charged black holes exhibit a suppressed Hawking temperature, particularly for extremal configurations, leading to a global reduction in flux. However, the angular profile remains qualitatively similar, though less intense overall.

The net effect is a highly anisotropic neutrino emission profile that is strongly collimated along the PBH’s direction of motion but modulated in strength by both spin and charge. This beaming behavior has important implications for detection. If multiple PBHs with similar velocity vectors evaporate in the same region of the sky (e.g., originating from a clustered population), their combined neutrino emission may yield localized and directional bursts. Such signals, if observed by detectors with sufficient angular resolution, could provide a unique handle on both the existence and kinematics of the PBH population.

\subsection{Energy Spectrum and Total Flux}
\label{subsec:energy_spectrum}

In addition to the angular anisotropy discussed earlier, the energy spectrum of neutrinos emitted from evaporating PBHs is strongly influenced by the PBH's velocity, spin, and charge. In the rest frame of the PBH, the differential neutrino emission spectrum follows a quasi-thermal distribution, modified by greybody factors that encode the spacetime's barrier potential for spin-1/2 particles. The general expression is given by:
\begin{align}
\left. \frac{d^2N}{dE\,dt} \right|_{\text{rest}} \propto \sum_{\ell, m} \frac{\Gamma_{\ell m s}(E, a, Q)}{2\pi} \cdot \left|{}_sS_{\ell m}(\theta, aE)\right|^2 \cdot \left[ \frac{1}{\exp(E/T_H) + 1} \right],
\end{align}
where \( \Gamma_{\ell m s}(E, a, Q) \) is the greybody factor for the spin-\( s \) field (with \( s = 1/2 \) for neutrinos), \( {}_sS_{\ell m}(\theta, aE) \) denotes the spin-weighted spheroidal harmonics, and \( T_H \) is the Hawking temperature, which itself depends on the PBH mass \( M \), spin \( a \), and charge \( Q \). 

When transformed to the lab frame, the energy spectrum experiences relativistic Doppler shifting and aberration due to the PBH motion. The transformation leads to a significant hardening of the spectrum for highly boosted PBHs (\( \beta \sim 0.9 \)), resulting in enhanced high-energy tails that may exceed detector thresholds even for sub-TeV Hawking temperatures.

\vspace{1em}
\begin{figure}[h!]
  \centering
  \begin{minipage}[b]{0.4\textwidth}
    \includegraphics[width=\textwidth]{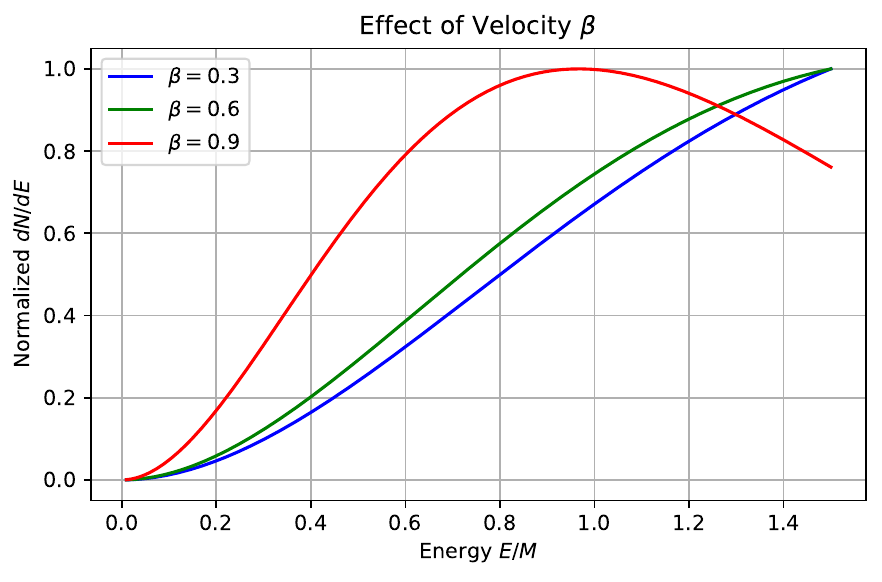}
  \end{minipage}
  \begin{minipage}[b]{0.4\textwidth}
    \includegraphics[width=\textwidth]{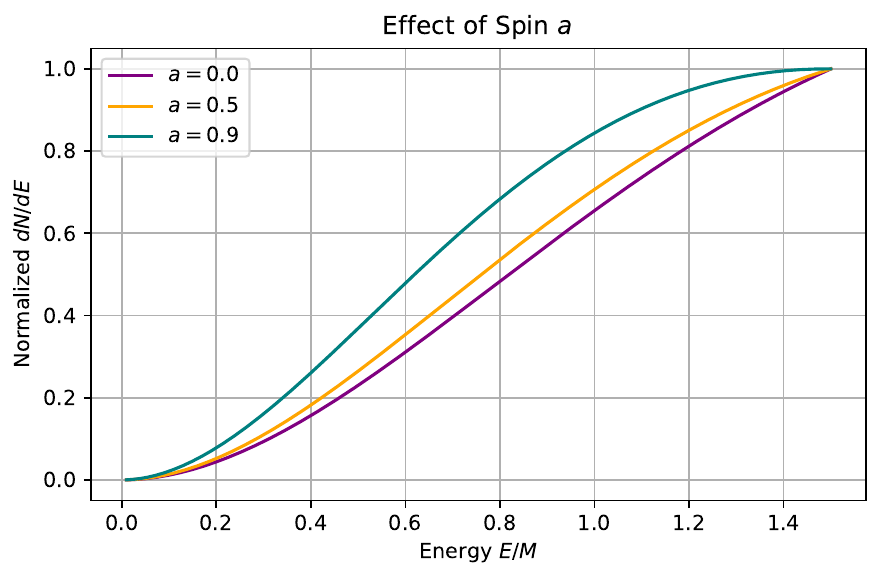}
  \end{minipage}
  \begin{minipage}[b]{0.5\textwidth}
    \includegraphics[width=\textwidth]{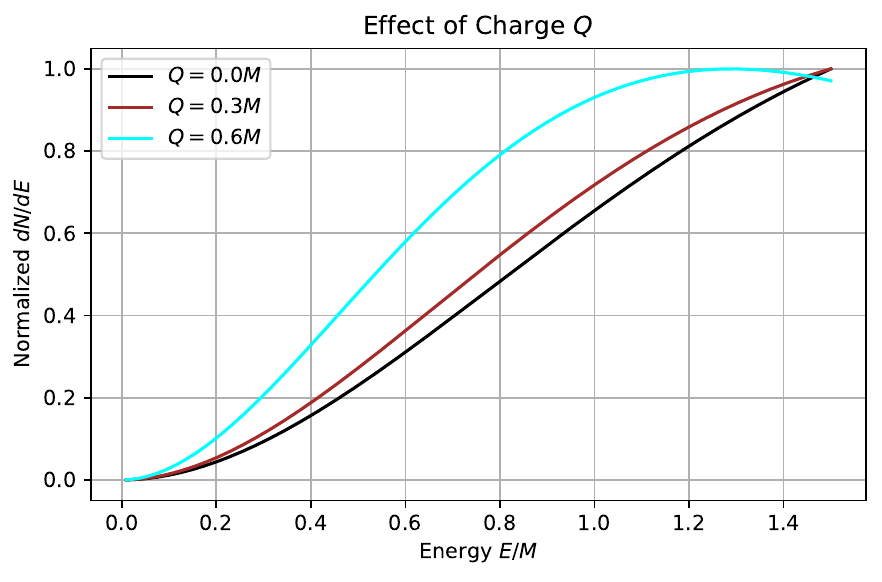}
  \end{minipage}
  \caption{Lab-frame neutrino spectra \( dN/dE \) for PBHs with varying physical parameters. Left: effect of increasing velocity \( \beta = 0.3, 0.6, 0.9 \) at fixed \( a = Q = 0 \). Right: variation with spin \( a = 0, 0.5, 0.9 \) at fixed \( \beta = 0.6 \), \( Q = 0 \). Bottom: impact of charge \( Q = 0, 0.3M, 0.6M \) at fixed \( \beta = 0.6 \), \( a = 0 \). All spectra are for  (\( M_{BH} = 10^{12} \))g.}
 \label{fig:energy_spectra}
\end{figure}
\vspace{1em}

Figure~\ref{fig:energy_spectra} shows the resulting lab-frame spectra for various values of velocity \( \beta \) (left), spin \( a \) (middle), and charge \( Q \) (right). Several salient features emerge:

\begin{itemize}
    \item \textbf{Velocity dependence:} As \( \beta \) increases, the energy spectrum is increasingly boosted in the forward direction, enhancing the high-energy tail. This is a direct consequence of Lorentz boosting and results in a larger fraction of neutrinos with energies above typical detection thresholds.
    
    \item \textbf{Spin effects:} Higher values of spin \( a \) introduce superradiant modes, particularly for particles with angular momentum aligned with the spin axis. These modes amplify low-energy emission and broaden the spectral profile, although the overall hardening is modest compared to velocity-induced boosts.
    
    \item \textbf{Charge effects:} As seen in the right panel, increasing the PBH charge \( Q \) suppresses the overall spectrum. This suppression originates from a lower Hawking temperature due to increased surface gravity at the horizon and reduced black hole evaporation rate. Although neutrinos are neutral and not directly affected by electromagnetic interactions, the reduction in temperature reduces the emission rate at all energies.
\end{itemize}

The bottom panel of Figure~\ref{fig:energy_spectra} thus reveals an important nuance: while spin \( a \) and velocity \( \beta \) enhance spectral hardness and overall flux, a sizable charge \( Q \) can suppress both, posing challenges for detection of PBHs with significant electric charge.

Finally, the detectability of these neutrino signals hinges not only on the integrated flux and angular concentration, but also on the spectral shape. Detectors such as IceCube or Hyper-Kamiokande have finite energy thresholds, and the hardening induced by relativistic motion can make otherwise undetectable Hawking neutrinos observable. Hence, understanding the lab-frame spectral modifications is crucial for identifying viable detection channels of evaporating PBHs.

\section{Detection Prospects and Constraints}
\label{sec:detection}

The detection of neutrinos emitted by evaporating PBHs is subject to several physical and instrumental constraints. In this section, we evaluate the prospects for detecting such neutrinos with existing and future detectors, focusing on both transient bursts and diffuse flux scenarios.

\subsection{Expected Event Rates}

We consider the lab-frame neutrino flux \( \phi(E, \theta) \) from a PBH of mass \( M \), Lorentz boost \( \beta \), and angular-dependent emission profile. The differential event rate in a detector of fiducial volume \( V \) is given by
\begin{align}
\frac{dN_{\text{events}}}{dt} = N_{\text{targets}} \int dE\, d\Omega\, \phi(E, \theta)\, \sigma_{\nu}(E)\, \epsilon(E),
\end{align}
where \( \sigma_{\nu}(E) \) is the neutrino interaction cross section, \( \epsilon(E) \) is the detector efficiency, and \( N_{\text{targets}} \) is the number of target particles (e.g., protons for inverse beta decay or argon nuclei for DUNE).

For concreteness, we take three benchmark detectors:
\begin{itemize}
    \item Super-Kamiokande (SK): water Cherenkov, \( V = 22.5\, \text{kt} \), sensitive to \( \bar{\nu}_e + p \to e^+ + n \)
    \item DUNE: liquid argon TPC, \( V = 40\, \text{kt} \), sensitive to \( \nu_e + ^{40}\text{Ar} \to e^- + X \)
    \item Hyper-Kamiokande (HK): next-generation water Cherenkov, \( V = 190\, \text{kt} \)
\end{itemize}

The cross sections used are:
\begin{align}
\sigma_{\bar{\nu}_e p}(E) &\approx 9.5 \times 10^{-44} \left( \frac{E}{\text{MeV}} \right)^2\, \text{cm}^2, \\
\sigma_{\nu_e \text{Ar}}(E) &\approx 10^{-38} \left( \frac{E}{\text{GeV}} \right)\, \text{cm}^2.
\end{align}

\begin{figure}[t]
\centering
\includegraphics[width=0.65\textwidth]{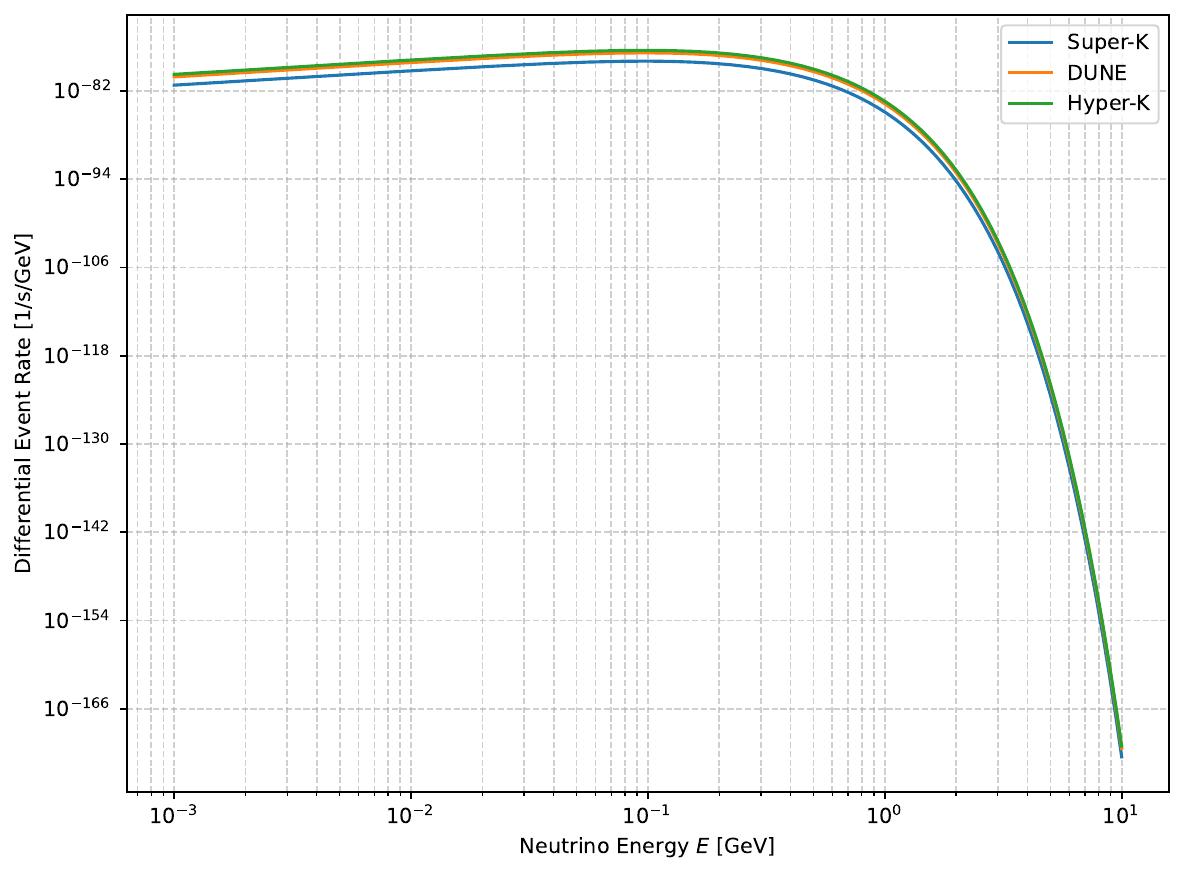}
\caption{Estimated number of neutrino events from a single evaporating PBH at \( D = 1\, \text{kpc} \) for Super-K, DUNE, and Hyper-K. We assume \( \beta = 0.9 \), unit absorption efficiency, and integrate over the full lab-frame spectrum.}
\label{fig:eventrate_distance}
\end{figure}

We estimate the total number of events for a single PBH at a distance \( D = 1\, \text{kpc} \) with mass \( M = 10^{12}\, \text{g} \), assuming instantaneous burst emission. Figure~\ref{fig:eventrate_distance} shows the event rates as a function of PBH mass for the three detectors. The signal peaks for PBH masses around \( 10^{11} \text{--} 10^{14} \, \text{g} \), where the neutrino spectrum overlaps with detector sensitivity and evaporation occurs quickly enough to be treated as a burst.

\subsection{Directional Signatures and Anisotropy}

The lab-frame emission is highly anisotropic due to relativistic beaming, especially for \( \beta \gtrsim 0.5 \). This leads to a forward-peaked neutrino burst aligned with the PBH motion. Such directionality can be exploited in detectors with angular resolution capabilities, such as water Cherenkov and liquid scintillator detectors.

\begin{figure}[t]
\centering
\includegraphics[width=0.65\textwidth]{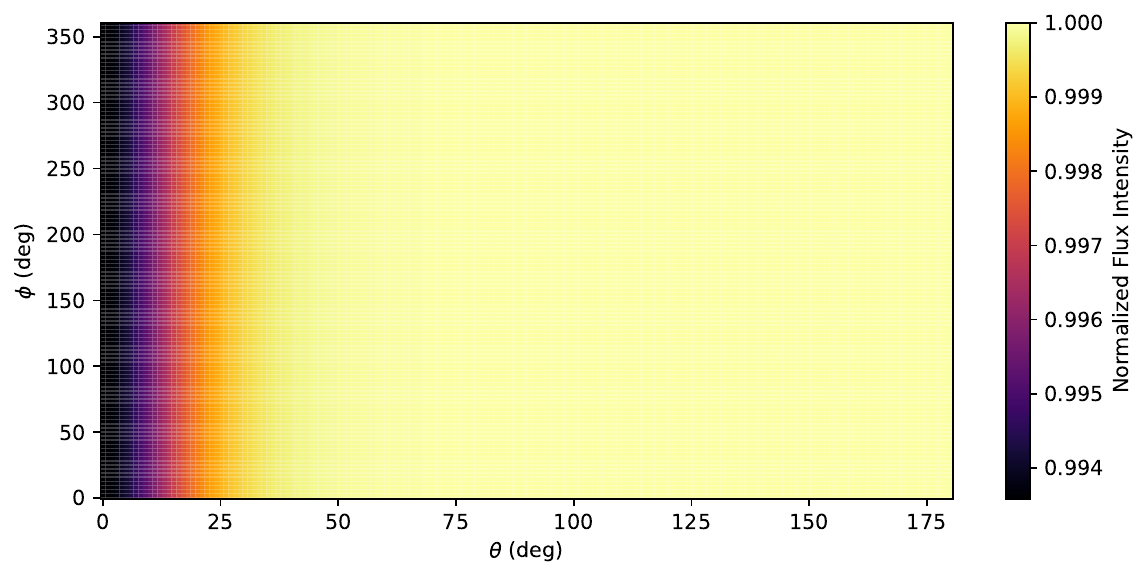}
\caption{Sky map of normalized neutrino flux intensity in lab frame for a PBH with \( \beta = 0.9 \), showing strong forward-beaming along direction of motion. The color scale represents \( \phi(E, \theta) \) integrated over energy.}
\label{fig:directional_map}
\end{figure}

Figure~\ref{fig:directional_map} illustrates the angular intensity distribution on the sky for a PBH with \( \beta = 0.9 \), showing a concentrated flux within a few degrees of the forward axis. If multiple PBHs contribute from similar directions—e.g., clustered within a dark matter subhalo—an excess over the diffuse neutrino background may be detectable as an anisotropic feature.

\subsection{Diffuse Background and Transient Bursts}

The observed signal depends on the spatial and temporal distribution of PBHs. We consider two scenarios:
\begin{itemize}
    \item Diffuse: numerous PBHs evaporating across cosmic history, contributing to a steady isotropic neutrino flux.
    \item Burst: local PBH evaporation events (e.g., within the Milky Way halo), producing transient signals detectable as individual bursts.
\end{itemize}

\begin{figure}[h!]
  \centering
  \begin{minipage}[b]{0.5\textwidth}
    \includegraphics[width=\textwidth]{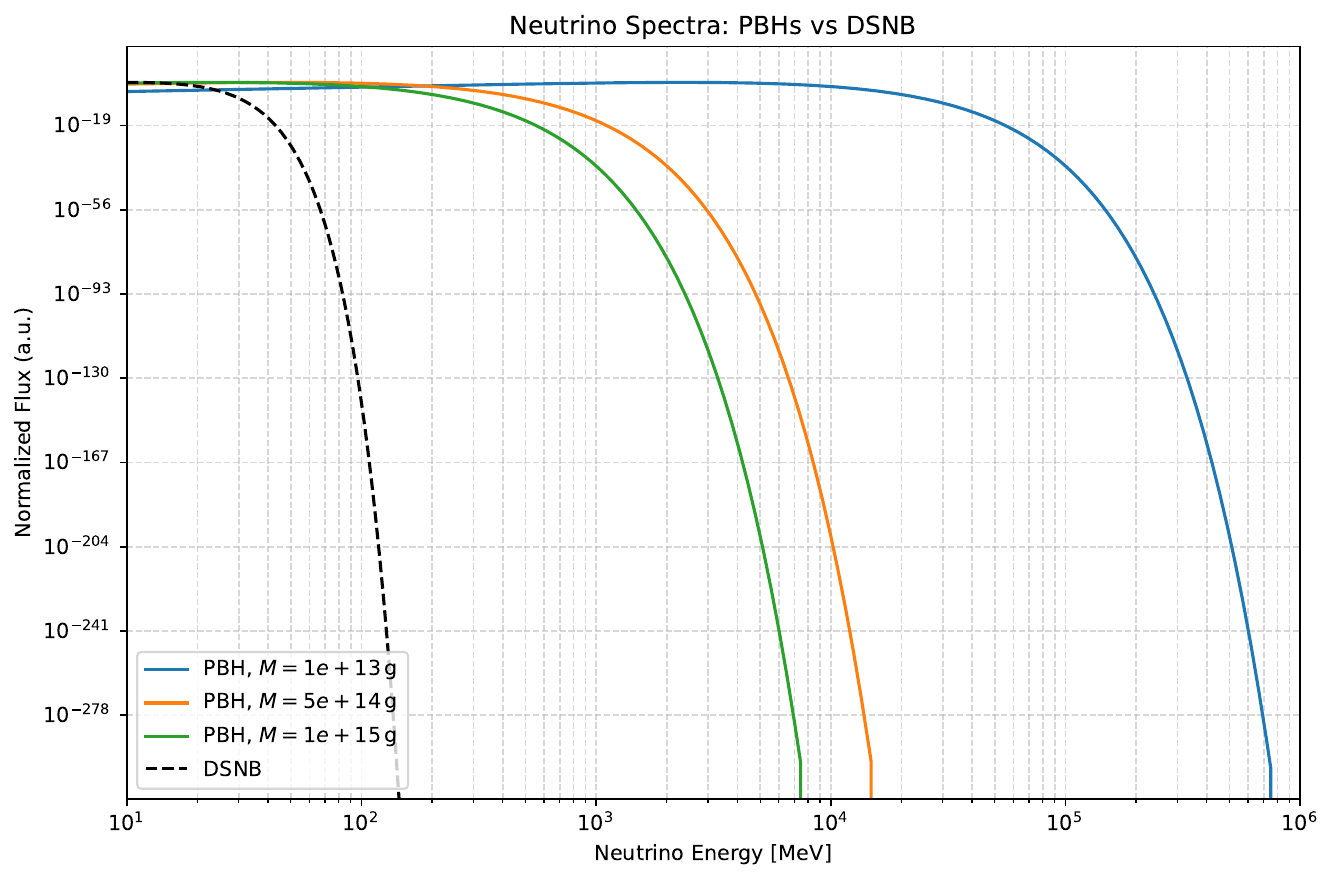}
  \end{minipage}
  \begin{minipage}[b]{0.5\textwidth}
    \includegraphics[width=\textwidth]{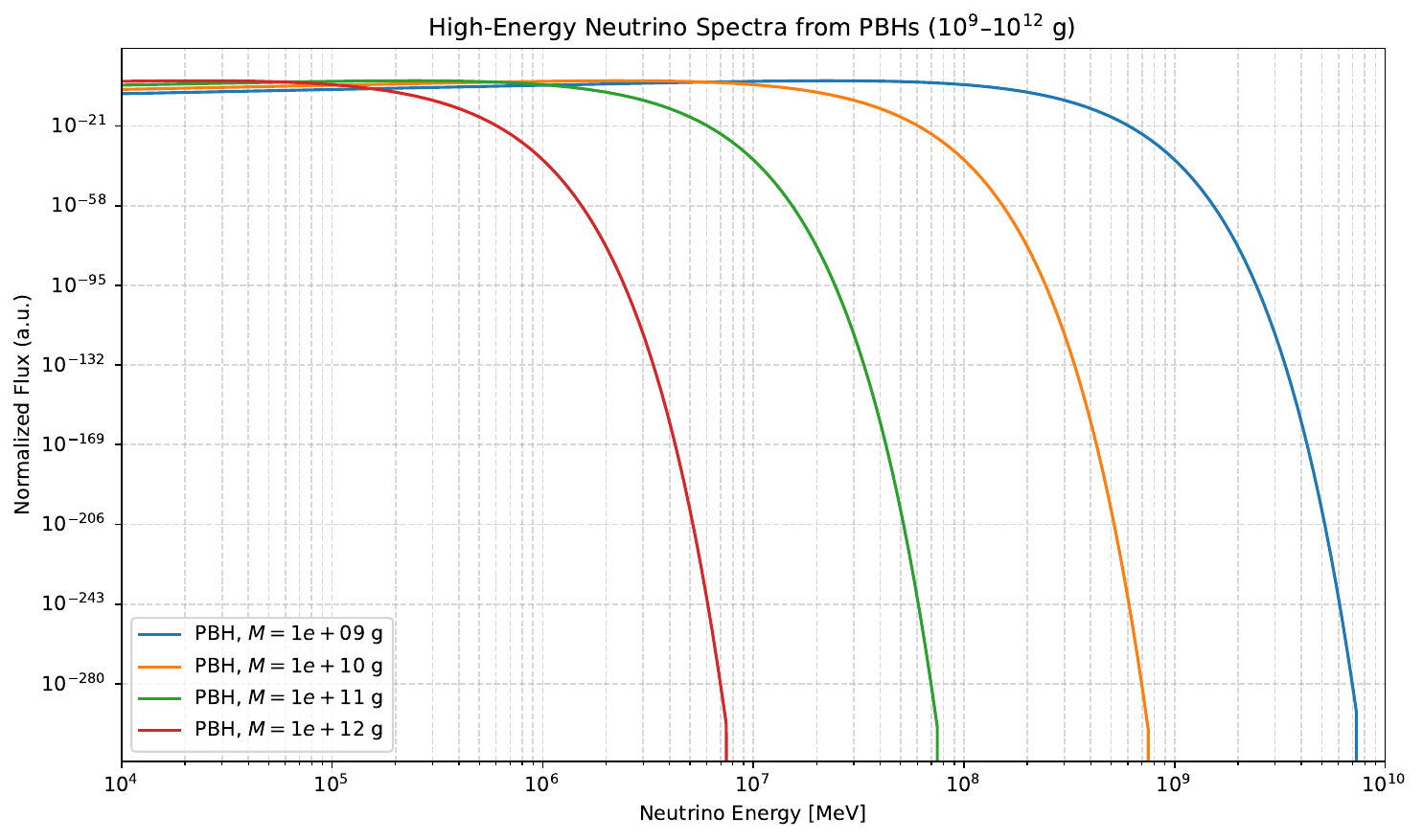}
  \end{minipage}
  \caption{Comparison of PBH neutrino flux (solid curves for different \( M \)) with DSNB (dashed) fluxes (top panel). PBHs with lower mass emit harder spectra as seen from the bottom panel.}
 \label{fig:energy_spectra1}
\end{figure}

The diffuse flux must compete with the diffuse supernova neutrino background (DSNB), which has an expected flux of \( \sim 10 \, \text{cm}^{-2}\text{s}^{-1} \) in the 10--50 MeV range. Figure~\ref{fig:energy_spectra1} compares the PBH neutrino flux to the DSNB backgrounds, demonstrating that PBHs can dominate at higher energies (\( E \gtrsim 100 \, \text{MeV} \)) if the local PBH density is sufficiently high.

\vspace{1em}
\vspace{1em}

\subsection{Constraints from BBN and DSNB}

The evaporation of PBHs before or during Big Bang Nucleosynthesis (BBN) injects high-energy particles that can dissociate light nuclei, placing stringent bounds on their abundance \cite{Kohri:1999ex, Carr:2009jm}. Similarly, an overproduction of neutrinos would distort the DSNB spectrum or exceed observed limits.

\begin{figure}[t]
\centering
\includegraphics[width=0.65\textwidth]{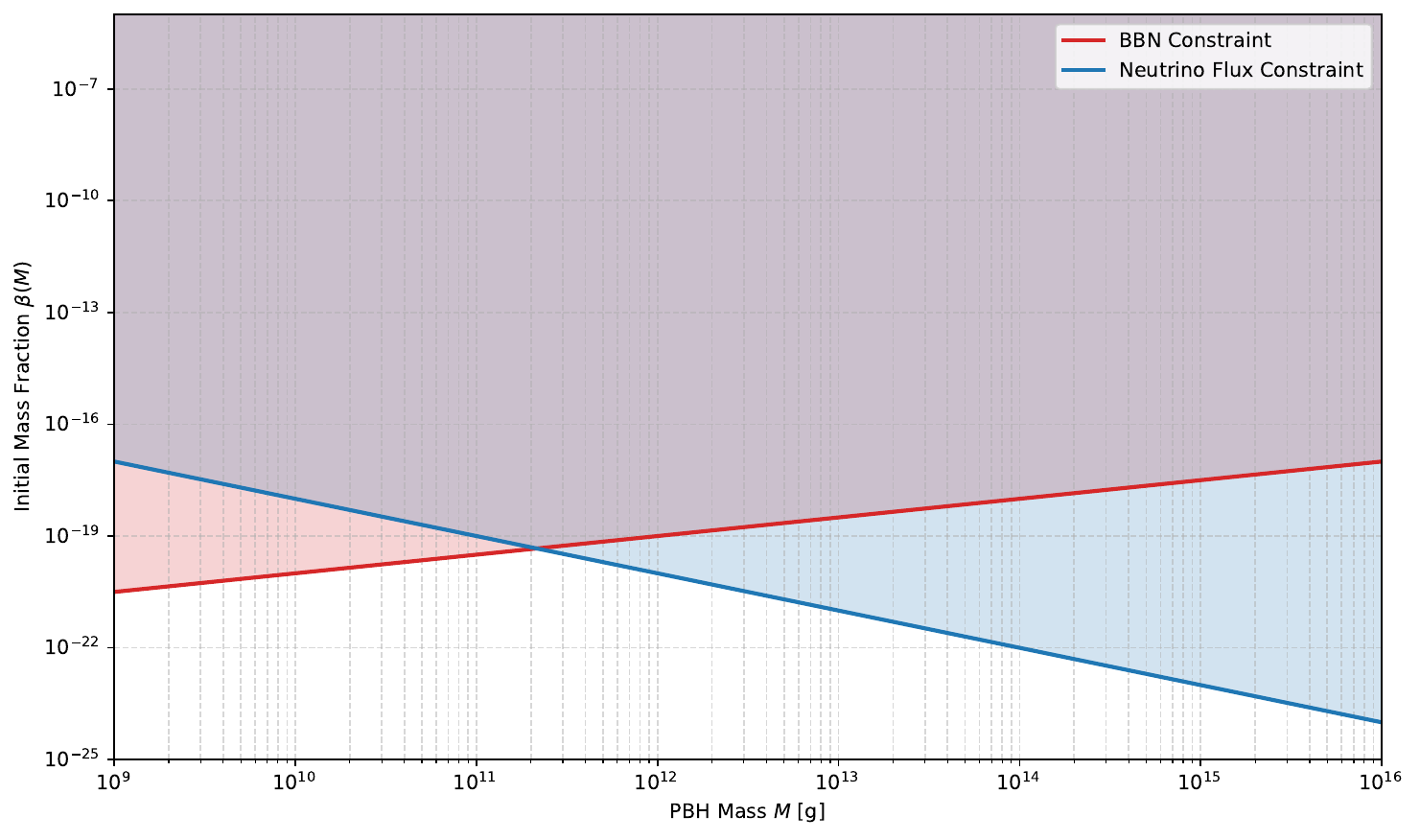}
\caption{Constraints on the initial PBH mass fraction \( \beta(M) \) from BBN and extragalactic neutrino flux limits. Regions above the curves are excluded. Our benchmark PBH masses are indicated for reference.}
\label{fig:bbn_constraint}
\end{figure}

To remain compatible with current cosmological data, the PBH mass spectrum must avoid excessive contribution in the range \( M \lesssim 10^9\, \text{g} \), unless the PBH population is highly suppressed or their formation is confined to late-time scenarios (e.g., phase transitions). Figure~\ref{fig:bbn_constraint} shows the allowed regions in the PBH abundance–mass plane.

\section{Cosmological Implications and Broader Outlook}
\label{cosmo}

The anisotropic emission of high-energy neutrinos from charged, spinning and moving PBHs opens several new avenues for probing the physics of the early Universe. The combined effects of spin-induced angular anisotropy and Doppler boosting imply that evaporating PBHs can act as transient, highly directional neutrino sources, producing bursts whose detectability depends sensitively on their redshift distribution, mass spectrum, and kinematic properties.

From a cosmological perspective, the detection of such bursts would provide direct information about PBH formation scenarios, including the small-scale features of the primordial power spectrum and possible non-Gaussianities that enhance collapse probabilities. Since PBHs can form over a wide mass range, constraints derived from directional neutrino signals are complementary to those from gravitational wave searches, microlensing surveys, and gamma-ray background measurements. In particular, they can extend sensitivity to low-mass PBHs whose evaporation occurs after recombination but before the present epoch, a regime that is challenging for other probes.

Directional neutrino bursts may also have implications for the cosmic neutrino background (C$\nu$B) anisotropy at ultra-high energies. While the average C$\nu$B is isotropic, a small subset of PBHs with relativistic bulk motion could imprint localized, transient anisotropies at energies far above the thermal scale. Additionally, such bursts could contribute to the ultra-high-energy neutrino population relevant for multimessenger astrophysics, offering cross-correlation opportunities with gamma-ray and cosmic-ray observatories.

Future high-statistics neutrino detectors with good angular resolution—such as IceCube-Gen2, GRAND, and radio-based neutrino arrays—will be capable of identifying the temporal and spatial clustering patterns expected from these events. This opens the possibility of constraining the fraction of dark matter in PBHs, the PBH spin distribution, and their typical velocities at late times. These measurements could be combined with gravitational wave observations from PBH binary mergers to produce a more complete cosmological census of PBHs.

Beyond PBHs themselves, the observation (or absence) of such directional bursts could also inform broader questions in cosmology, such as the history of small-scale structure formation, the nature of primordial perturbations, and the potential existence of exotic particle species influencing the evaporation spectrum. In this way, anisotropic neutrino emission from PBHs acts as a new and highly targeted probe of the early Universe, complementary to established cosmological and astrophysical messengers.

\section{Conclusion}
\label{sec:conclusion}

In this work, we have investigated in detail the angular, spectral, and temporal properties of neutrinos emitted by primordial black holes (PBHs) that simultaneously carry spin, electric charge, and relativistic motion. By incorporating the Kerr--Newman geometry, spin-weighted spheroidal harmonics, and Lorentz transformations, we have shown that the interplay of these effects produces distinctive and highly directional features in the lab-frame neutrino flux that cannot be captured by the isotropic Schwarzschild approximation.

We have demonstrated that rotation induces nontrivial angular structures in the rest-frame emission through the combined action of spin-dependent greybody factors and mode enhancement for positive azimuthal numbers. Bulk motion then compresses and blueshifts this anisotropic radiation into a narrow forward cone, substantially increasing the flux and hardening the spectrum along the direction of motion. The addition of electric charge modifies the Hawking temperature and greybody factors, indirectly affecting the neutrino channel by altering the available energy budget and the evaporation timescale. For sufficiently large charge-to-mass ratios, this suppression of temperature leads to softer spectra and reduced fluence, while also prolonging the lifetime of the PBH.

%Our event-rate calculations for current and upcoming neutrino detectors indicate that such directional beaming can significantly improve detectability, especially for PBHs with masses around \(10^{9}\)–\(10^{10}\,\mathrm{g}\) evaporating in the present epoch with high boosts. The results further show that the combination of spin, motion, and charge imprints a unique angular morphology and spectral hardness that could serve as a powerful discriminant against other astrophysical sources. Cosmological constraints from Big Bang Nucleosynthesis and the diffuse supernova neutrino background remain relevant, but they are modified when the evaporation history is shaped by nonzero spin and charge.

These findings highlight a novel prediction: evaporating Kerr--Newman PBHs in motion can produce short, highly directional neutrino bursts whose properties encode detailed information about the PBH mass, spin, charge, and velocity. This opens a new observational window complementary to gamma-ray, microlensing, and gravitational-wave probes. Looking forward, several developments are needed to bring this scenario closer to observational tests: accurate numerical greybody factors for spin-1/2 fields in Kerr--Newman spacetimes, realistic PBH population and velocity models from formation scenarios, full detector-response folding for directional burst searches, and exploration of flavor oscillation, helicity asymmetries, and beyond–Standard-Model emission channels. Coordination with gamma-ray and gravitational-wave facilities could further enhance the discovery potential by enabling multi-messenger confirmation of candidate events.

In conclusion, the combined effects of spin, charge, and relativistic motion qualitatively enrich the phenomenology of Hawking evaporation, with neutrinos offering a uniquely clean probe of these processes. Their unattenuated propagation and sensitivity to horizon-scale physics make them an ideal channel for exploring the small-scale structure of the early Universe. With improved theoretical modeling and dedicated experimental searches, directional neutrino observations could become a decisive tool for testing the existence and properties of low-mass PBHs.

\acknowledgments
AC acknowledges the financial support of the Japan Society for the Promotion of Science (JSPS) through the JSPS Postdoctoral Fellowship (Standard), grant number JP23KF0289. AC also extends sincere gratitude to Prof. Baradhwaj Coleppa for his invaluable feedback, which significantly contributed to the refinement of this work.

\appendix
\section{Greybody Factor}
\label{app:greybody}

The greybody factor encodes the probability that a particle produced near the black hole horizon will escape to infinity, modifying the purely thermal spectrum predicted by Hawking’s original calculation. For Kerr–Newman black holes, this factor depends nontrivially on the particle’s spin \( s \), energy \( \omega \), angular momentum quantum numbers \( (\ell, m) \), the black hole’s mass \( M \), spin parameter \( a \), and charge \( Q \). It is obtained by solving the appropriate Teukolsky equation with ingoing boundary conditions at the horizon and purely outgoing conditions at spatial infinity.

For spin-\(\frac{1}{2}\) particles such as neutrinos, the relevant radial Teukolsky equation in Boyer–Lindquist coordinates can be written as
\begin{equation}
\Delta^{-s} \frac{d}{dr} \left( \Delta^{s+1} \frac{dR_{\ell m \omega}}{dr} \right) + \left[ \frac{K^2 - 2is(r - M)K}{\Delta} + 4is\omega r - \lambda_{\ell m \omega} \right] R_{\ell m \omega} = 0 ,
\label{eq:teukolsky_radial}
\end{equation}
where \( \Delta = r^2 - 2Mr + a^2 + Q^2 \), \( K = (r^2 + a^2)\omega - am - qQ r \) with \( q \) the particle’s electric charge (vanishing for neutrinos), and \( \lambda_{\ell m \omega} \) is the angular separation constant obtained from the spin-weighted spheroidal harmonics \( {}_sS_{\ell m}(\theta; a\omega) \). The greybody factor for a given mode is defined as
\begin{equation}
\Gamma_{\ell m}^s(\omega; a, Q) \equiv 1 - \left| \frac{\mathcal{R}_{\ell m \omega}}{\mathcal{I}_{\ell m \omega}} \right|^2 ,
\label{eq:greybody_def}
\end{equation}
where \( \mathcal{I}_{\ell m \omega} \) and \( \mathcal{R}_{\ell m \omega} \) are the ingoing and reflected wave amplitudes at infinity. For massless fermions, \(\Gamma_{\ell m}^s\) rises from zero at low energies to unity at high energies, with a frequency-dependent structure reflecting superradiant suppression (for bosons) or enhancement effects absent in the fermionic case.

In the low-frequency limit \( \omega M \ll 1 \), analytic expressions can be obtained by matching near-horizon and far-field solutions~\cite{Page:1976df, Page:1977um}. For neutrinos from a neutral Kerr black hole (\(Q=0\)), the dominant \(\ell = 1/2\) mode has a greybody factor scaling as
\begin{equation}
\Gamma_{1/2, m}^{1/2}(\omega) \simeq 4 (\omega r_+)^2 + \mathcal{O}\!\left( (\omega r_+)^3 \right),
\end{equation}
where \( r_+ \) is the outer horizon radius. Inclusion of nonzero \( Q \) modifies both the potential barrier height and the horizon radius, thus altering this scaling. While neutrinos are unaffected by the Coulomb term \( qQ/r \), the change in the metric functions modifies their transmission probability.

For spinning and moving PBHs, the greybody factor enters the boosted lab-frame spectrum via
\begin{equation}
\frac{d^3 N}{dtd\omega d\Omega} = \sum_{\ell, m} \frac{\Gamma_{\ell m}^s(\omega; a, Q)}{\exp\!\left[ \frac{\omega - m\Omega_H}{T_H} \right] + 1} \, \left| {}_sS_{\ell m}(\theta'; a\omega') \right|^2 ,
\end{equation}
where \( \omega' \) and \( \theta' \) are the rest-frame frequency and angle related to lab-frame variables by a Lorentz transformation, \( \Omega_H \) is the horizon angular velocity, and \( T_H \) is the Hawking temperature for Kerr–Newman geometry. The angular dependence from \( {}_sS_{\ell m} \) combines with the boost-induced aberration to shape the observed directional profile.

Although we have used semi-analytic approximations for \(\Gamma_{\ell m}^s\) in this work, a fully accurate computation for Kerr–Newman backgrounds with arbitrary \( a \) and \( Q \) requires numerical integration of Eq.~\eqref{eq:teukolsky_radial} and mode-by-mode matching, which we leave for future study. Such calculations will be essential for precision predictions of PBH-induced neutrino signals, especially when seeking to extract black hole parameters from directional burst observations.

\end{document}